\documentclass[aps,showpacs,floatfix,twocolumn,prb,superscriptaddress]{revtex4-1}
\usepackage{amssymb,amsthm, amssymb, latexsym}
\usepackage{graphicx}
\usepackage[utf8]{inputenc}
\usepackage{amsmath}
\usepackage[english]{babel}
\usepackage{float}
 \usepackage{color}

\begin{document}

\title{Haldane phase in the sawtooth lattice: Edge states,
  entanglement spectrum and the flat band}

\author{Beno\^{i}t Gr\'{e}maud}
\affiliation{MajuLab, CNRS-UNS-NUS-NTU International
Joint Research Unit UMI 3654, Singapore}
\affiliation{Centre for Quantum Technologies, National
University of Singapore, 2 Science Drive 3, 117542 Singapore}
\affiliation{Department of Physics, National University of
Singapore, 2 Science Drive 3, 117542 Singapore}
\affiliation{Laboratoire Kastler Brossel, UPMC-Sorbonne Universit\'es,
CNRS, ENS-PSL Research University, Coll\`{e}ge de France, 4 Place
Jussieu, 75005 Paris, France}

\author{G. George Batrouni}
\affiliation{Universit\'e C{\^o}te d'Azur, INLN, CNRS, France}
\affiliation{MajuLab, CNRS-UNS-NUS-NTU International
Joint Research Unit UMI 3654, Singapore}
\affiliation{Centre for Quantum Technologies, National
University of Singapore, 2 Science Drive 3, 117542 Singapore}

\begin{abstract}

  Using density matrix renormalization group numerical calculations,
  we study the phase diagram of the half filled Bose-Hubbard system in
  the sawtooth lattice with strong frustration in the kinetic energy
  term. We focus in particular on values of the hopping terms which
  produce a flat band and show that, in the presence of contact and
  near neighbor repulsion, three phases exist: Mott insulator (MI),
  charge density wave (CDW), and the topological Haldane insulating
  (HI) phase which displays edge states and particle imbalance between
  the two ends of the system. We find that, even though the
  entanglement spectrum in the Haldane phase is not doubly degenerate,
  it is in excellent agreement with the entanglement spectrum of the
  Affleck-Kennedy-Lieb-Tasaki (AKLT) state built in the Wannier basis
  associated with the flat band.  This emphasizes that the absence of
  degeneracy in the entanglement spectrum is not necessarily a
  signature of a non-topological phase, but rather that the (hidden)
  protecting symmetry involves non-local states.  Finally, we also
  show that the HI phase is stable against small departure from
  flatness of the band but is destroyed for larger ones.

\end{abstract}


\maketitle

\section{Introduction}
\label{intro}
Since its introduction by Fisher {\it et al.} \cite{fisher89}, the
bosonic Hubbard model (BHM) and its variants have attracted a great
deal of attention due to the rich physics and wide variety of phases
and phase transitions it exhibits: incompressible Mott insulator (MI),
superfluid (SF), Bose glass in the presence of disorder
\cite{fisher89}; charge density wave (CDW), supersolid (SS) and phase
separation in the presence of longer range repulsion
\cite{batrouni14,batrouni95,batrouni00,goral02,wessel05,boninsegni05,sengupta05,otterlo05,batrouni06,yi07,suzuki07,dang08,pollet10,capogrosso10}. Interest
intensified when trapped atomic condensates where loaded in optical
lattices \cite{greiner02} where it was shown that the system is
governed by the BHM \cite{jaksch99}. The high tunability of the
interaction strengths, the wide range of lattice geometries that can
be realized and the ability to perform very detailed measurements
offered access to a very wide range systems and tight binding
Hamiltonians of great interest in the study of strongly correlated
systems.

Increasingly, over the last several years, the physics of strongly
correlated quantum systems has focused on the existence and properties
of unconventional phases and phase transitions. For example, when the
extended one-dimensional Bose Hubbard model (EBH) is at full filling
and the contact interaction dominates, the system is in the MI phase
with one particle per site.  As the near neighbor interaction is
increased, quantum fluctuations create holes and doublons, typically
leading to three kinds of sites: empty, singly occupied and doubly
occupied. One can then approximate the model by the spin-$1$
Heisenberg chain with $S_z=1, 0, -1$ for the doubly, singly and empty
sites.  This led to the discovery \cite{altman06,altman08} that at
full filling, the one dimensional extended BHM supports, in addition
to the SF, MI and CDW phases, a topological phase characterized by a
nonlocal string order parameter and edge states. This phase, called
the Haldane insulator (HI), is closely related to the Haldane phase
\cite{haldane83} appearing in integer spin chains and realized in the
AKLT state \cite{aklt87,kt92}. The phase diagram of this model was
elaborated and the excitation spectra studied using quantum Monte
Carlo (QMC) simulations and density matrix renormalization group
(DMRG) calculations \cite{rossini12,ggb13,ggb14,ggb16}. Another
situation where unconventional phases may be encountered is that of
strong geometrical frustration, for example in the particle kinetic
energy. If the hopping term in the Hamiltonian is frustrated in a
particular way, the lowest band can become flat resulting in a huge
degeneracy of states into which the particles may condense. This
degeneracy changes the behavior of the particles when their density is
sufficiently low (or when there is no interaction between them): The
particles are localized in extended geometrical structures due to
interference effects of the hopping terms. What happens when
interaction is present and the particle filling exceeds a critical
value dictated by the lattice geometry was treated by several
authors\cite{huber10,takayoshi13,tovmasyan13}. A general approach used
in these references is to project the Hamiltonian onto the flat band
resulting in an effective low energy Hamiltonian which depends only on
the interaction. For example, in Ref.[\onlinecite{huber10}], the
sawtooth and the kagome lattices were treated in this way. For the
sawtooth lattice, the effective projected Hamiltonian describes bosons
hopping on a one-dimensional lattice and interacting with each
other. The question then arises if, under certain situations, this
model can support a topological phase, such as the HI, as happens in
the extended one-dimensional BHM mentioned above.

This is the main question which we will address in this work. We will
show, with DMRG calculations, that by adding an extended interaction
term to the model studied in Ref.[\onlinecite{huber10}], and when the
system has a flat band and is at half filling, the charge and neutral
gaps behave in a way reminiscent of the HI. We then show the presence
of edge states and imbalance in the number of particles on the two
ends of the system (with open boundary conditions) which confirms the
existence of the HI. We also show that, contrary to the usual
situation, the entanglement spectrum does not exhibit a doubly
degenerate ground state in the Haldane phase. We then demonstrate that
this behavior persists in a finite interval of geometric frustration
around the flat band but disappears if the frustration is far from
what is needed for flatness.

The paper is organized as follows. In section \ref{modelsection} we
present the model and review the results of
Ref.~\onlinecite{huber10}. In section \ref{gapssection} we present our
DMRG results for the gaps and density profiles and discuss the various
phases found at half filling. Topological aspects of the edge states
are presented and discussed in section \ref{edgestatesection} while
the entanglement spectrum is discussed in section
\ref{wanniersection}. The phase diagrams when the lowest band is
dispersive but still almost flat are presented in section
\ref{notflatsection} followed by our conclusions in section
\ref{conclusionsection}.

\section{Model}
\label{modelsection}

\begin{figure}[h!]
\centerline{\includegraphics[width=8cm]{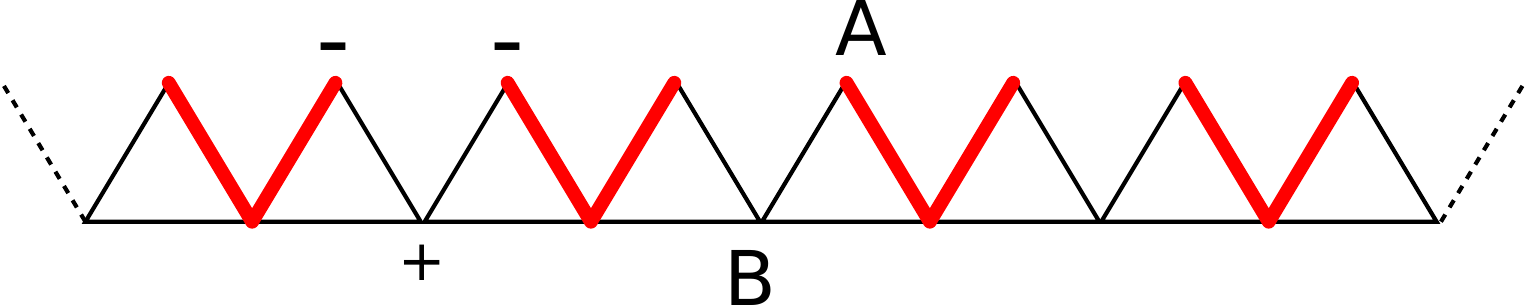}}
\caption{(Color online) The sawtooth lattice. Sites at the bases of
  the triangles are denoted $B$ and apex sites are $A$. The localized
  state wavefunction has positive amplitude on the $B$ sites and
  negative on the $A$ sites (see text). The thick (red) lines in the
  form of {\sf V} are the states localized in the flat band. The sites
  labeled $A$ and $B$ belong to the same unit cell.
  Our convention for the lattice names, i.e. AA, BB and AB, indicates
  the type of site at each end of the lattice with open boundary
  conditions. }
\label{sawtoothlattice}
\end{figure}

We will study the BHM on the sawtooth lattice,
Fig.\ref{sawtoothlattice}, and governed by the Hamiltonian

\begin{eqnarray}
\label{sawtoothham}
\nonumber
H &=&\sum_{\langle ij\rangle} |t_{ij}|
\left ( a_{i}^{\dagger}a_{j}^{\phantom\dagger} + {\rm H.c}\right )
+\frac{U}{2}\sum_{i}n_i(n_i-1)\\
&&+V\sum_{\langle ij\rangle_B} n_i n_j,
\end{eqnarray}
where $\langle ij\rangle$ denotes near neighbors and $\langle
ij\rangle_B$ near neighbors of the $B$ type; the operators $a_i$ and
$a_i^\dagger$ are destruction and creation operators on site $i$ and
satisfy the usual bosonic commutation relations. $n_i$ is the number
operator on site $i$ and $t_{ij}$ is the hopping parameter between
near neighbor sites. We will take $t_{ij}=t$ for hops between two B
sites (see Fig.\ref{sawtoothlattice}) and $t_{ij}=t^{\prime}$ for hops
between A and B sites. The contact interaction strength is $U$ and the
near neighbor interaction is $V$ both of which are taken to be
repulsive. Note that the near neighbor repulsion is only between
neighboring $B$ sites for reasons to be discussed below.

The kinetic part of the Hamiltonian, Eq.(\ref{sawtoothham}), can be
easily diagonalized\cite{huber10} giving the band energies
\begin{equation}
\epsilon_{\pm}(k)=\cos (ka)\pm\sqrt{t^2 \cos^2(ka)+2 t'^2 \cos (ka)+2t'^{2}}.
\end{equation}
When $t^{\prime}=\sqrt{2}t$, the lowest band becomes flat and we get,
\begin{align}
\epsilon_{\pm}(k)=
\begin{cases}
\quad\phantom{-}2t\left (1+\cos(ka)\right ),\\
\quad-2t,
\end{cases}
\end{align}
where $a$ is the distance between two $B$ sites.

If a particle is now introduced in the system, it will be localized on
three sites, in the shape of {\sf V}, due to the interference effects
of frustration. These sites are shown in thick (red) {\sf V}-shaped
structures in Fig.\ref{sawtoothlattice}. The localized states are
given by,
\begin{equation}
\label{vshape}
|V_i\rangle = \frac{1}{2} \left ( \sqrt{2}
  a_{B,i}^{\dagger}-a_{A,i-1}^{\dagger}-a_{A,i}^{\dagger} \right )
|0\rangle,
\end{equation}
where, on average, a $B$ site is occupied by $1/2$ a particle and each
of the two $A$ sites by $1/4$. It is clear from
Fig.\ref{sawtoothlattice} that even in the presence of interaction, if
the total number of particles is less than or equal to $1/4$ of the
number of sites, these localized states are eigenstates of $H$. If the
particle density exceeds the critical density of $1/4$, the {\sf V}
states will start sharing $A$ sites and, therefore, interacting. The
energy per particle will no longer be $-2t$. Figure
\ref{sawtoothlattice} shows the configuration of {\sf V}'s for the
case of maximum filling without interacting and exhibits a CDW of
{\sf V}'s alternately empty and occupied by one particle.

The states $|V_i\rangle$ are linearly independent and complete in the
flat band subspace but they are not orthogonal. For this reason, we
use Wannier states to define a set of localized orthogonal states
forming a complete basis. In general, denoting by $w_{n\,X_p}(x)$ the
Wannier function for the band $n$ localized around the position $X_p$,
the Wannier states $|w_{n\,X_p}\rangle$ give rise to an orthogonal
basis if and only if the centers $X_p$ are on a Bravais lattice. In
that case, one can express the creation operators $a^{\dagger}_i$ as
functions of the creation operator $W^{\dagger}_{n\,X_p}$,
i.e. creating a boson in the Wannier state $|w_{n\,X_p}\rangle$:
\begin{equation}
\label{localtowannier}
a^{\dagger}_{i}=\sum_{n\,X_p}w^*_{n\,X_p}(x_i)W^{\dagger}_{n\,X_p},
\end{equation}
where $x_i$ is the position of the site $i$. The projection of the
operators on a given band $n$ is made by keeping, in the preceding
sum, the Wannier operators for that band only:
$a^{\dagger}_{i}\approx\sum_{X_p}w^*_{n\,X_p}(x_i)W^{\dagger}_{n\,X_p}$.

\begin{figure}[h!]
\centerline{\includegraphics[width=8cm]{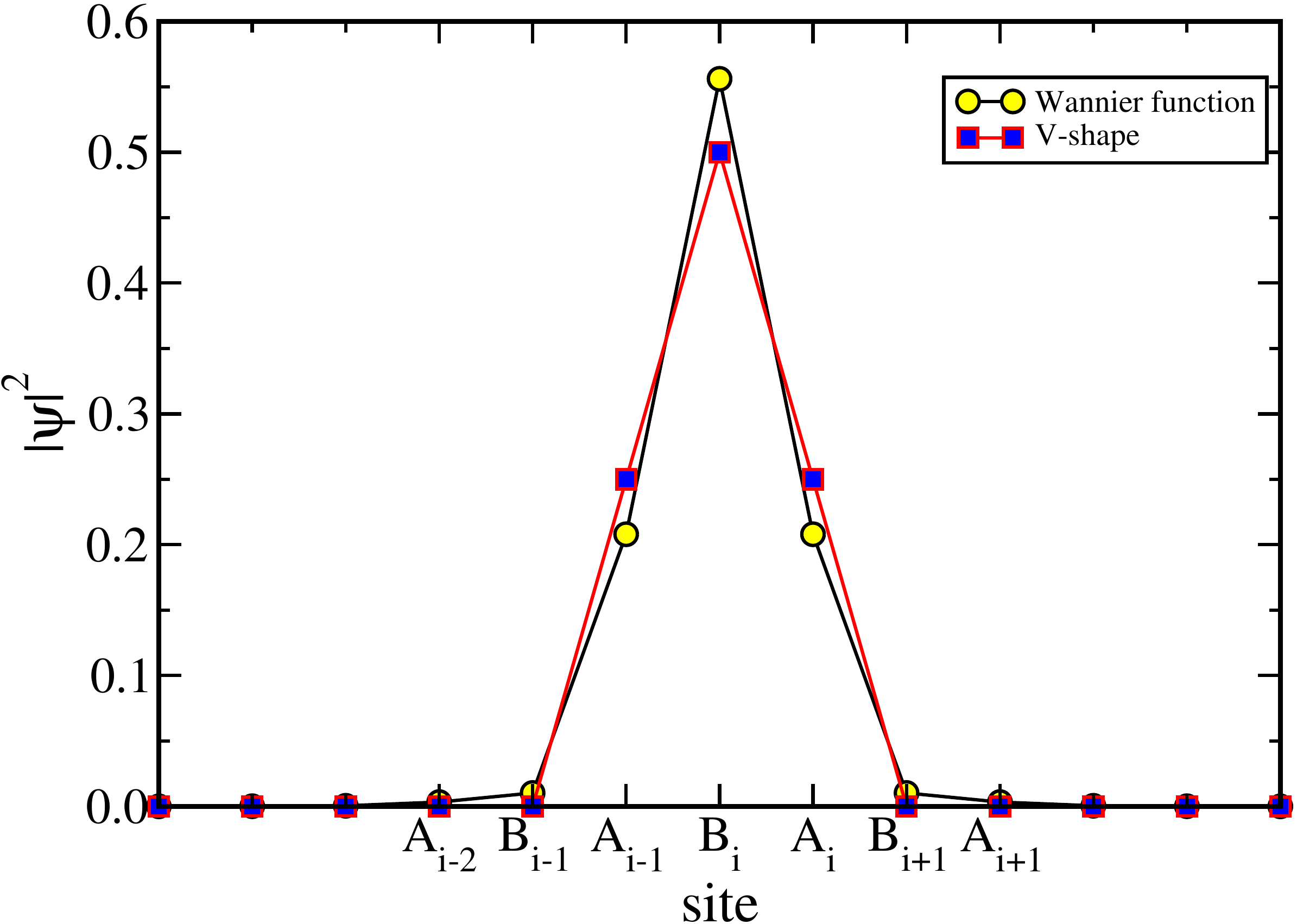}}
\caption{(Color online) Density profile of the Wannier function, see
  Eq.~\eqref{wfb}, and of the {\sf V}-shape, see
  Eq.~\eqref{vshape}. centered around the site $B_i$. The Wannier
  function is normalized to one, with a total density probability 0.58
  on the $B$ sites and 0.42 on the $A$ sites.}
\label{wannier_den}
\end{figure}

In the present case, the projection on the flat band leads to
\begin{eqnarray}
\label{abfb}
\nonumber
a^{\dagger}_{Bi}&\approx&\sum_{j}f_B^*(j-i)W^{\dagger}_{f\,j}\\
a^{\dagger}_{Ai}&\approx&\sum_{j}f_A^*(j-(i+1/2))W^{\dagger}_{f\,j},
\end{eqnarray}
where the Wannier state $|w^{\dagger}_{f\,j}\rangle$ has been chosen
to be localized around the $B$ site at position $ja$.  Later, we will also
need the expression of the Wannier operators as a function of the operators
$a^{\dagger}_{Bi}$ and $a^{\dagger}_{Ai}$:
\begin{equation}
\label{wfb}
 W^{\dagger}_{f\,j}=\sum_i\left(f_B(j-i)a^{\dagger}_{Bi}+f_A(j-(i+1/2))a^{\dagger}_{Ai}\right).
\end{equation}
From the band
structure, one can derive the amplitude of the Wannier function at the
$B$ and $A$ sites:
\begin{equation}
\begin{aligned}
  f_B^*(p)&= -\frac{2}{\pi} \int_{0}^{+\pi/2}d\theta\,
  \frac{\cos(2p\theta)} {\sqrt{1+2\cos^2(\theta)}}\\
  f_A^*((p-1/2))&= \frac{2}{\pi} \int_{0}^{+\pi/2}d\theta\,
  \frac{\sqrt{2}\cos((2p-1)\theta)\cos(\theta)}{\sqrt{1+2\cos^2(\theta)}}.
\end{aligned}
\end{equation}
The Wannier function on the flat band is plotted in
Fig.~\ref{wannier_den}. Note that our definition of these Wannier
states differs somewhat from that of reference [\onlinecite{huber10}].

Using DMRG calculations on the original and the effective
Hamiltonians, Ref.[\onlinecite{huber10}] showed, for $V=0$, that when
the system is doped slightly above the critical density of $1/4$, the
CDW melts yielding a SF with a momentum distribution peaking at a
nonzero value which depends on the doping.

Here, we are interested in the phases and phase transitions of this
system at half filling when the near neighbor repulsion, $V$, is
included. Specifically, the questions we address here are: Will the
half filled system admit a MI phase of {\sf V}'s sharing $A$ sites?
Will the system admit a CDW phase with alternating vacant and doubly
occupied {\sf V}'s? And, will the system admit a HI phase sandwiched
in between these two?

\section{DMRG: Phase diagram at half filling}
\label{gapssection}

To address the above questions, we use the DMRG codes in the
ALPS\cite{alps} library to calculate density profiles and the neutral
and charge energy gaps at half filling. We take $t=1$ to fix the
energy scale and focus attention on two values of the contact
interaction, $U=1/2, 1$, and study the system as the near neighbor
repulsion, $V$, is varied. Most of our results are for the flat band
case, $t^{\prime}=\sqrt{2} t$. The $t^{\prime} \neq \sqrt{2} t$ will
be addressed in section \ref{notflatsection}. We perform the
calculations for several system sizes, $L=40, 80, 120, 160, 200$
(number of sites is $2L$) with open boundary conditions and a maximum
number of boson per state $N_{max}=3$. The neutral gap, $\Delta_n$, is
calculated by targeting the ground and the first excited states in the
DMRG calculation. We typically keep $300$ states in the DMRG
calculation and we have verified that keeping more (up to $600$
states) or increasing $N_{max}$ does not change the results. The
charge gaps, $\Delta_c$, are calculated using the energies at half
filling and at half filling doped with one particle and doped with one
hole,
\begin{equation}
\Delta_c = E(N_{1/2}+1)+E(N_{1/2}-1)-2E(N_{1/2}),
\end{equation}
where $N_{1/2}$ is the number of particles at half filling.

Since we use open boundary conditions, our convention for the lattice
names, i.e. AA, BB and AB, indicates the type of site at each end of
the lattice and since the lattice is frustrated, special care must be
taken in order to observe the various phases. More precisely, in the
following, we will show that, at half filling, the system exhibits three types of insulating
phases: Mott, Haldane, and CDW phases.  In order to compute the gap
values, one needs to be able to lift the degeneracy among the
different ground states. As for the EBH on a 1D chain, this is done by
adding a chemical potential at each edge, but, since the ``natural''
local states after projection on the flat band are the {\sf V}-shaped
structures (or the Wannier function), the boundary conditions must be
matched in terms of these local states. In particular, adding a large
positive chemical potential on a $B$ site at one edge (and on the
neighboring $A$ site, if the lattice ends with an $A$ site) more or
less amounts to imposing a vanishing density on the {\sf V}-shape
localized on that $B$ site; on the other hand, adding a large negative
chemical potential on either ending $B$ (and on the neighboring $A$)
site does not correspond to getting a fixed number of bosons in the
{\sf V}-shape localized on this $B$ site. In other words, we do not
have a simple way to impose on the ground state to be in a Fock state
in the {\sf V}-shape (or Wannier) basis at the edge. This is in a
sharp contrast with the EBH on a 1D-chain, for which a large negative
chemical potential on a given site amounts to fixing the state on that
site to a Fock state with $N_{max}$ bosons.

From that point of view, our strategy was to compare the results for
the densities, gap values and correlation functions for different
lattice types, i.e. AA, BB and AB and different boundary conditions,
i.e. adding or not a large chemical potential at each edge. In what
follows, we adopt the following conventions: a lattice size $L$ always
means a lattice with $(L+1)$ $B$-sites, an AA lattice has then $(L+2)$
$A$-sites whereas a BB lattice has $L$ $A$-sites. Then, it turns out
that, for the different phases, the proper choice of the lattice type,
boundary conditions and the number of bosons that give the correct
charge and neutral gaps is are follows:
  
\begin{itemize}

\item[$\bullet$] MI phase (single ground state): AA lattice with $L+1$
  bosons and no additional chemical potential.
  
\item[$\bullet$] CDW insulating phase (two degenerate ground states):
  BB lattice with $L$ bosons, $L$ even and one large additional
  chemical potential on both the first and last B sites.

\item[$\bullet$] HI phase (four degenerate ground states): BB lattice
  with $L$ bosons and one large additional chemical potential on both
  the first and last B sites.

\end{itemize}
We confirmed that the preceding situations yield correct values for
the gaps by comparing with the gap values obtained with DMRG, for
$L=40$, under periodic boundary conditions where the above issues do
not arise. Larger sizes with periodic boundary conditions do not
converge properly with DMRG.

\begin{figure}[h!]
\centerline{\includegraphics[width=9cm]{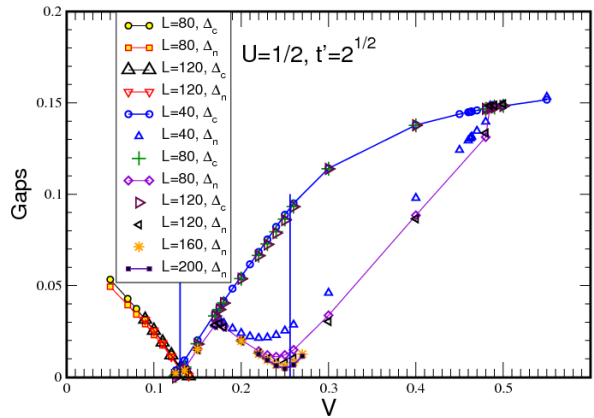}}
\caption{(Color online) DMRG results for the charge, $\Delta_c$, and
  neutral, $\Delta_n$, gaps as functions of the near neighbor
  repulsion, $V$, for $U=1/2$, $t^{\prime}=\sqrt{2}t$ (flat band) at
  half filling. The vertical lines show the locations of the MI-HI
  ($V\approx 0.13$) and the HI-CDW ($V=0.256$) transitions (see
  text).}
\label{gapsU0.5}
\end{figure}

\begin{figure}[h!]
\centerline{\includegraphics[width=9cm]{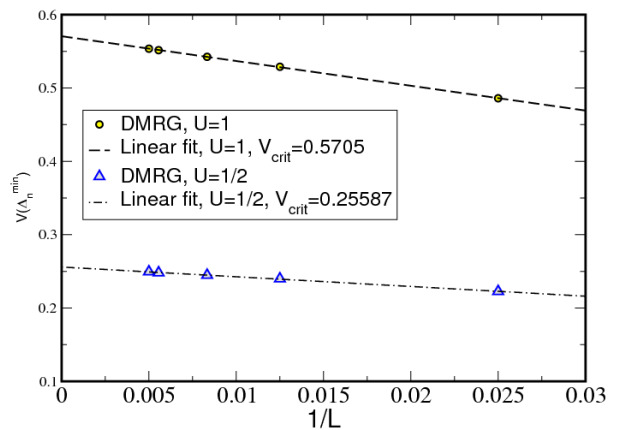}}
\caption{(Color online) The location of the minimum of $\Delta_n$ is
  sensitive to finite size effects. The value of $V$ where $\Delta_n$
  reaches its minimum is extrapolated to $1/L\to 0$ to obtain the
  HI-CDW transition value.}
\label{gapsextrap}
\end{figure}

\begin{figure}[h!]
\centerline{\includegraphics[width=9cm]{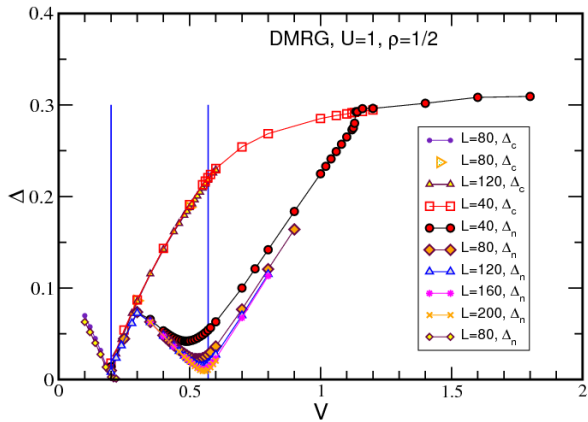}}
\caption{(Color online) DMRG results for the charge, $\Delta_c$, and
  neutral, $\Delta_n$, gaps as functions of the near neighbor
  repulsion, $V$, for $U=1$, $t^{\prime}=\sqrt{2}t$ (flat band) at
  half filling. The vertical lines show the locations of the MI-HI
  ($V\approx 0.2$) and the HI-CDW ($V=0.571$) transitions (see
  text). }
\label{gapsU1.0}
\end{figure}

Figure \ref{gapsU0.5} shows, for several lattice sizes, $\Delta_c$ and
$\Delta_n$ as functions of $V$ for $U=1/2$ and $t^{\prime}=\sqrt{2}$
at half filling. For $V<0.13$, $\Delta_c=\Delta_n$ and both gaps
vanish at $V\approx 0.13$ indicating a quantum phase transition at
that point. This behavior of the gaps is indicative of the
incompressible MI phase\cite{altman06}. For $V>0.13$, the gaps first
increase together, then at $V\approx 0.175$, $\Delta_n$ starts to
decrease while $\Delta_c$ continues increasing. $\Delta_n$ then
reaches another minimum, which approaches zero as $L$ increases, then
starts to increase again tending toward $\Delta_c$. This behavior of
$\Delta_n$ suggests the presence of the HI phase between its two
minima. The second minimum of $\Delta_n$ is sensitive to finite size
effects. We show in Fig.\ref{gapsextrap} how the critical value of $V$
is obtained by extrapolating the value of the minimum to $L\to
\infty$. This yields $V_{crit}=0.256$. Therefore, for $0.13 \leq V
\leq 0.256 $ the system is in the Haldane insulating phase. To confirm
that this is indeed a HI phase, one could calculate the string order
parameter\cite{altman06,altman08,rossini12,ggb13,ggb14,ggb16} by using
the Wannier operators as the ``site'' operators.  However, the
nonlocality of the Wannier states leads to extremely complicated
expressions in terms of the local operators $a_i$, which makes it
impractical to compute this Wannier-string order parameter.  Instead,
we will confirm below the nature of this HI phase by studying
topological properties such as the presence of edge states and the
entanglement spectrum. For $V>0.256$ the system is in the CDW phase as
we will see by examining the density profiles below. Figure
\ref{gapsU1.0} shows similar behavior to Fig.\ref{gapsU0.5} but for
$U=1$.

\begin{figure}[h!]
\centerline{\includegraphics[width=9cm]{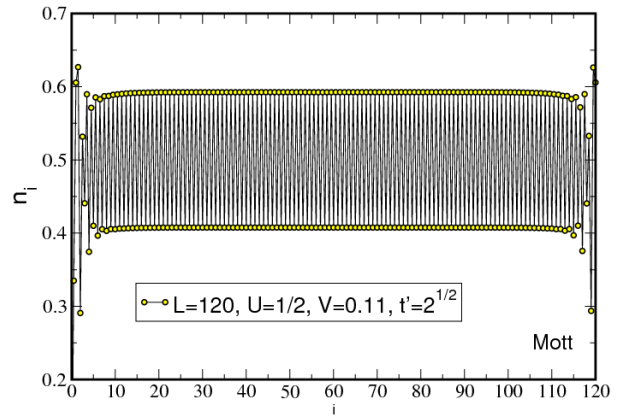}}
\caption{(Color online) Density profile in the MI phase. The $B$ sites
are labeled by integer valued site index $i$ while the $A$ sites are
labeled by half odd integer. Away from the boundaries, average $B$-site
occupation is $0.41$ while that of $A$-sites is $0.59$.}
\label{MIdenprof}
\end{figure}

The density profiles give additional information on the various
phases. In order to show the density profile of the sawtooth lattice
as a two-dimensional plot, we assigned integer (half odd integer)
labels to the $B$ ($A$) sites. Figure \ref{MIdenprof} shows the
density profile, $n_i$, in the MI phase for $U=1/2$ and $V=0.11$. We
see that in the bulk, {\it i.e.} away from the ends of the system, the
average filling of an $A$ site is $\langle n_A \rangle =0.59$ while
for a $B$ site it is $\langle n_B \rangle =0.41$. As $V$ is decreased,
$\langle n_A \rangle$ and $\langle n_B \rangle$ approach $1/2$. This
is easy to understand if one recalls the basic localized structure
discussed above. In such a localized state, the $B$-site has an
average occupation $1/2$ and the $A$-site has $1/4$. At half filling
(the case we are considering here) the system is full of touching {\sf
  V}'s, so now the average $A$ filling is $1/2$ too. The near neighbor
repulsion between $B$ sites disfavors $B$ occupation thus pushing
$\langle n_B \rangle $ down while the value of $\langle n_A \rangle$
increases. This will be discussed further in section
\ref{wanniersection}.

\begin{figure}[h!]
\centerline{\includegraphics[width=9cm]{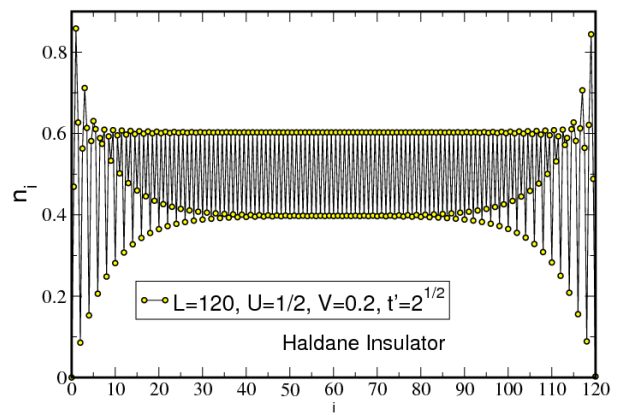}}
\caption{(Color online) The density profile in the HI phase. In the
  bulk, away from the ends of the system, $\langle n_A\rangle =0.6$
  and $\langle n_B\rangle =0.4$. Near its ends, the system exhibits
  evidence of edge states whose penetration into the system depends on
  the value of $V$. See text.}
\label{HIdenprof}
\end{figure}

\begin{figure}[h!]
\centerline{\includegraphics[width=9cm]{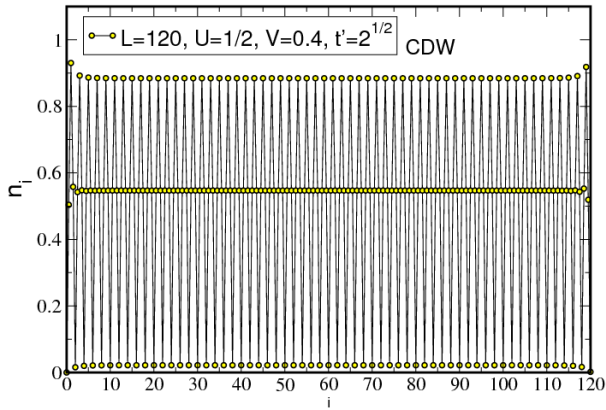}}
\caption{(Color online) The density profile in the CDW phase. $\langle
  n_A\rangle \approx 0.55$ while $\langle n_B\rangle$ alternates
  between almost empty sites, $\langle n_B\rangle \approx 0.022$, and
  $\langle n_B\rangle \approx 0.88$ indicating a CDW of {\sf V}
  structures (see Fig.\ref{sawtoothlattice}).}
\label{CDWdenprof}
\end{figure}

Figure \ref{HIdenprof} shows the density profile in the HI phase,
between the two minima of $\Delta_n$. In the bulk, away from the ends
of the system, $\langle n_A\rangle =0.6$ and $\langle n_B\rangle
=0.4$. Closer to the ends of the system we see evidence of edge states
both in the $B$-site and the $A$-site density profiles. In this
figure, the edge states extend around 40 sites into the system. This
penetration depends on $V$, it decreases as $V$ is taken closer to the
MI-HI transition and increases when $V$ is taken closer to the HI-CDW
transition. This will be discussed in more detail in sections
\ref{edgestatesection} and \ref{wanniersection}. The presence of the
edge states is added confirmation that the phase between the two
minima of $\Delta_n$ is indeed HI.

For $V>0.256$, the system is in the CDW phase: $A$-sites have a
constant occupation while $B$-sites alternate between high and low
filling. In Fig.\ref{CDWdenprof}, $V=0.4$, we have $\langle n_A\rangle
\approx 0.55$ and $\langle n_B\rangle$ alternates between almost empty
sites $\langle n_B\rangle \approx 0.022$ and $\langle n_B\rangle
\approx 0.88$. As $V$ increases, the occupation of the low-filling
$B$-sites decreases further. Roughly speaking, this can be understood
in terms alternating doubly occupied {\sf V}'s, in analogy with the
alternating doubly occupied sites in the CDW phase of the
one-dimensional BHM on a simple chain.

\begin{figure}[h!]
\centerline{\includegraphics[width=8cm]{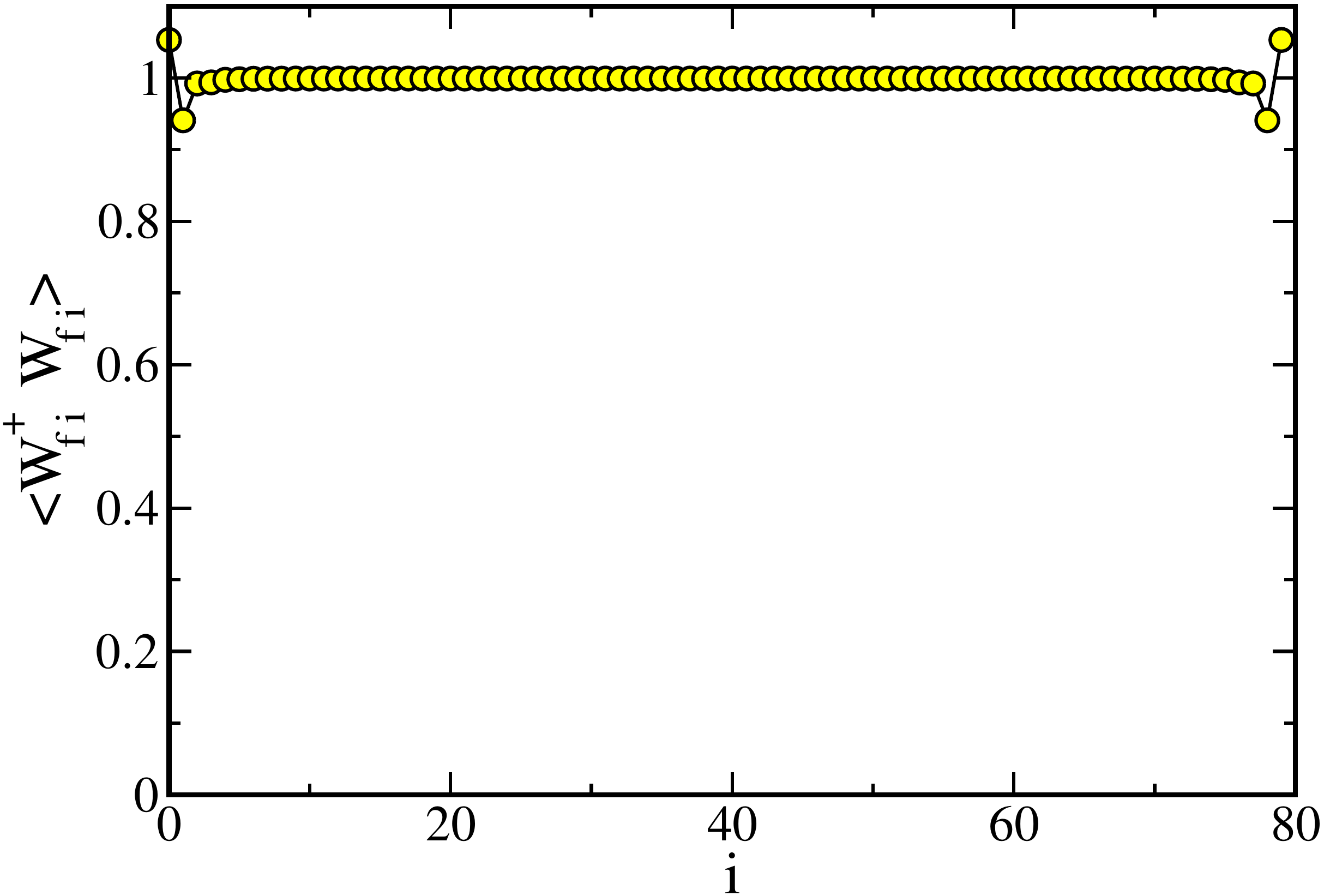}}
\caption{(Color online) Wannier basis density profile in the MI phase,
  $U=0.5$, $V=0.05$.  The average density is equal to 1. The small
  oscillations at both ends reflect the truncation of the Wannier
  function due to the boundary.}
\label{MIdenprof-wannier}
\end{figure}

\begin{figure}[h!]
\centerline{\includegraphics[width=8cm]{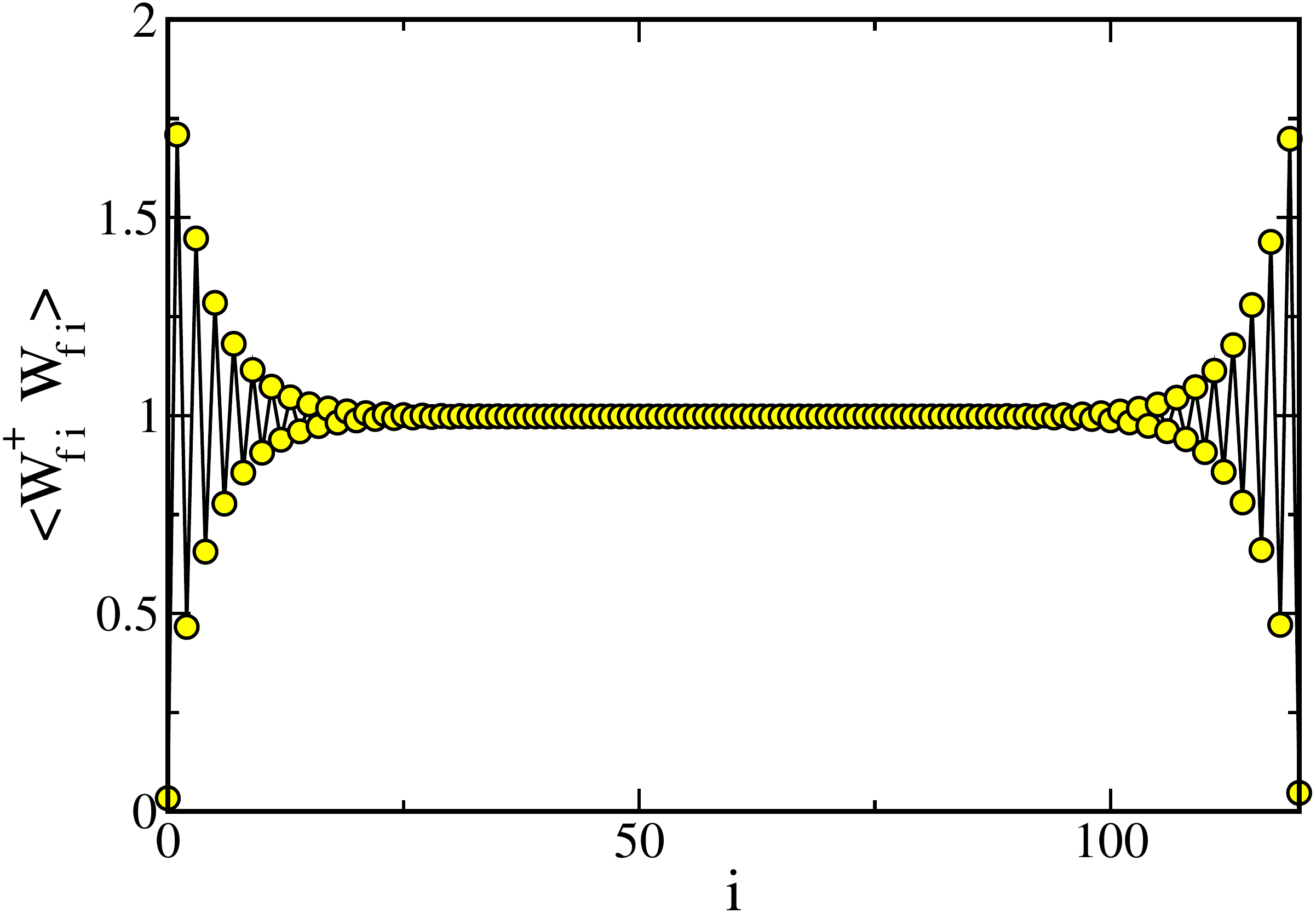}}
\caption{(Color online) Wannier basis density profile in the HI phase
  $U=0.5$, $V=0.17$. In the bulk, away from the ends of the system,
  the average density is one. Near its ends, the system exhibits clear
  evidence of edge states whose penetration into the system depends on
  the value of $V$. Due to the boundary conditions (see text), the
  effective charge of each edge state is $-1/2$.  }
\label{HIdenprof-wannier}
\end{figure}

\begin{figure}[h!]
\centerline{\includegraphics[width=8cm]{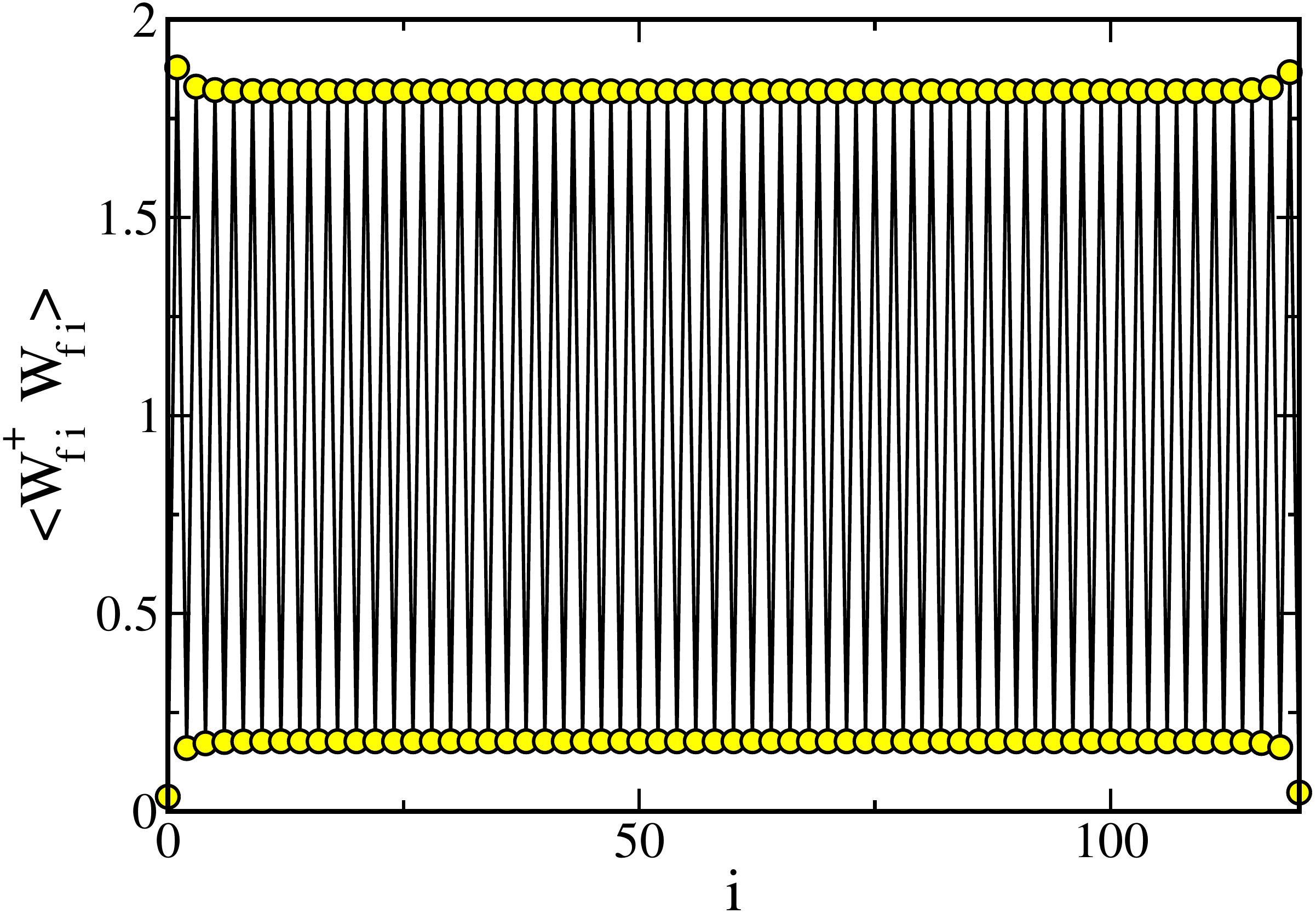}}
\caption{(Color online) Wannier basis density profile in the CDW phase
  $U=0.5$, $V=0.4$. The density alternates between almost empty sites
  ($\langle n_B\rangle \approx 0.18$ and $\langle n_B\rangle \approx
  1.82$ indicating a clear CDW in the Wannier states.}
\label{CDWdenprof-wannier}
\end{figure}

As explained above, the gap behavior and the density profiles are very
reminiscent of the MI-HI-CDW transitions in the usual EBH
model\cite{altman06,altman08,rossini12,ggb13,ggb14,ggb16}.  To
emphasize this point, we have used Eq.~\eqref{wfb} to compute the
average density in the Wannier states. More precisely, we have
computed $\langle W^{\dagger}_{f\,i}W^{\phantom\dagger}_{f\,i}\rangle$
where $i$ represents the center of the Wannier state, see
Fig.~\ref{wannier_den}. The corresponding density profiles in the MI,
HI and CDW phases are shown in
Figs.~\ref{MIdenprof-wannier},\ref{HIdenprof-wannier} and
\ref{CDWdenprof-wannier} respectively. As expected, in the MI the
density profile is flat with unit filling, whereas in the HI, the
profile exhibits nice edge states at the boundaries and a uniform
density in the bulk (unit filling). On the contrary, in the CDW, the
density alternates between almost empty sites, $\langle n_B\rangle
\approx 0.18$ and $\langle n_B\rangle \approx 1.82$. Even though the
density profile looks very simple in the MI phase, it is not just
given by a naive mean-field ansatz consisting of the tensor product of
singly occupied Wannier functions:
\begin{equation}
\label{mimf}
|\mathrm{MI}\rangle=\otimes_{i=1}^L|1\,i\rangle,
\end{equation}
where $|n\,i\rangle$ is the Fock state having $n$ bosons in the
Wannier state centered around the $B$ site $i$. Indeed, using
Eq.~\eqref{wfb} one can compute the average local densities,
$\langle\mathrm{MI}|\hat{n}_A|\mathrm{MI}\rangle$ on $A$ sites and
$\langle\mathrm{MI}|\hat{n}_B|\mathrm{MI}\rangle$ on $B$ sites. This
yields $\langle n_A\rangle\approx0.42$ and $\langle
n_B\rangle\approx0.58$. These values actually correspond to the total
probability density on $A$ and $B$ sites of a single Wannier function and are very
different from the ones computed from the ground state, i.e. $\langle
n_A \rangle =0.59$ and $\langle n_B \rangle =0.41$.  This strong
discrepancy between the ground state of the system and the mean-field
ansatz given by Eq.~\eqref{mimf} emphasizes that the doublon-holon
states (in the Wannier basis) actually have a large contribution to
the ground state in the MI phase.

\section{Edge states}
\label{edgestatesection}

A feature of the topological Haldane phase, as typified by the AKLT
state, is the appearance of edge states when the system has open
boundaries. Depending on the boundary conditions, the edge states that
appear in the usual EBH on a 1D-chain can exhibit either an excess or
a deficit of half a boson compared to the total number of sites
occupied by the edge states.

In the sawtooth lattice case, calculating the number of particles on
the left and right halves of the system in Figs. \ref{MIdenprof} and
\ref{CDWdenprof} we find balanced populations and, therefore, no
topological effects in the MI and CDW phases. However, the situation
is different in the case of Fig. \ref{HIdenprof} where we find $59.8$
particles on the left side and $60.2$ on the right. We verified this
is not a finite size effect by doing the same calculations for $L=160,
200$ and for several $V$ values and also keeping more states in the
DMRG calculation. In all cases we found the same imbalance by $\pm0.2$
particles relative to half filling. This value of $0.2$ agrees very
well with $\langle n_B \rangle/2$, i.e. as if half the system loses
half the occupation of a single $B$ site which goes to the other half
of the system.

This becomes even clearer by examining the edge states directly in the
Wannier basis, i.e. using Eq.~\eqref{wfb} to compute $\langle
W^{\dagger}_{f\,i}W^{\phantom\dagger}_{f\,i}\rangle$ where $i$
represents the centre of the Wannier state as plotted in
Fig.~\ref{HIdenprof-wannier}, which resembles very closely the usual
EBH on a 1D chain. More precisely, as explained in
Sec.~\ref{modelsection}, the BB lattice in our simulations amounts to
imposing a vanishing density on both B-sites, which is then analogous
to the usual EBH on a 1D chain with vanishing densities at the end. In
that situation, the HI ground state is obtained when the number of
bosons is one less than the total number of sites, such that, after
splitting the system in two parts, each edge state contains a
half-boson less than the number of sites in their respective parts.
Indeed, in our case, the total number of Wannier ``sites'', is $121$,
but the total number of bosons $N=120$, because of the vanishing
densities at both boundaries. Actually, the total density in the
Wannier states is $119.74$ because of the truncation of the Wannier
function at the boundaries. When splitting the system in two parts,
say the left one with 61 sites and the right one with 60 sites, the
total number of bosons is 60.37 to the left and is 59.37 to the
right. As one can readily verify, the $0.74$ boson has been split in
two between the left and the right parts. Finally, since the average
density in the bulk is equal to unity, changing the splitting point
would only change these numbers by integer values without affecting
the fractional part. This analysis shows the HI phase which we have
found in our model, genuinely exhibits edge states with a fractional
filling.

\begin{figure}[h!]
\centerline{\includegraphics[width=9cm]{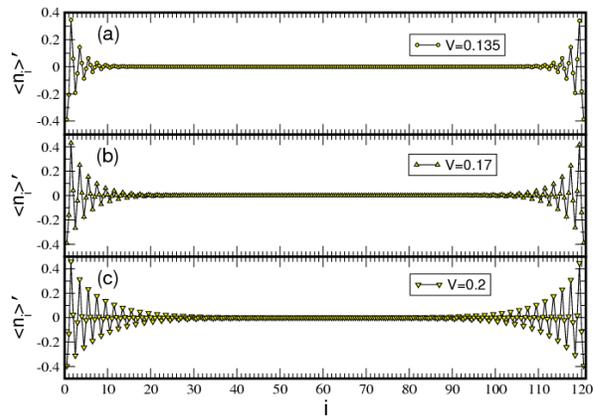}}
\caption{(Color online) The shifted density profiles, $\langle
  n_i\rangle^{\prime}$, in the HI phase (see text) for $L=120$,
$U=1/2$. The extent of the edge state into the system increases as $V$
approaches the HI-CDW transition value. The envelope is exponential.}
\label{HIdenprof-shifted}
\end{figure}

In order to bring out more clearly these edge states in our DMRG
results, we subtract the bulk values of $\langle n_A \rangle $ and
$\langle n_B \rangle $ from the occupations of the $A$ and $B$ sites
in the density profiles in the HI phase. In other words, for all
integer sites ({\it i.e.} $B$-sites) we subtract the bulk value of
$\langle n_B\rangle $, and for all half odd integer sites ({\it i.e.}
$A$-sites) we subtract the bulk value of $\langle n_A\rangle $. This
yields the shifted density profiles in
Fig.\ref{HIdenprof-shifted}. The figure, which resembles
Fig.~\ref{HIdenprof-wannier}, shows that close to the MI-HI transition
($V=0.135$) the edge states extend a few sites into the system on each
side. As $V$ increases and the system approaches the HI-CDW
transition, the edge states extension increases until, for a fixed
system size, $L$, the two edge states start to overlap. When this
happens, a larger system is needed to get precise results. When the
HI-CDW transition point is reached, $\Delta_n$ vanishes and the system
makes the transition to the CDW phase.

\begin{figure}[h!]
\centerline{\includegraphics[width=9cm]{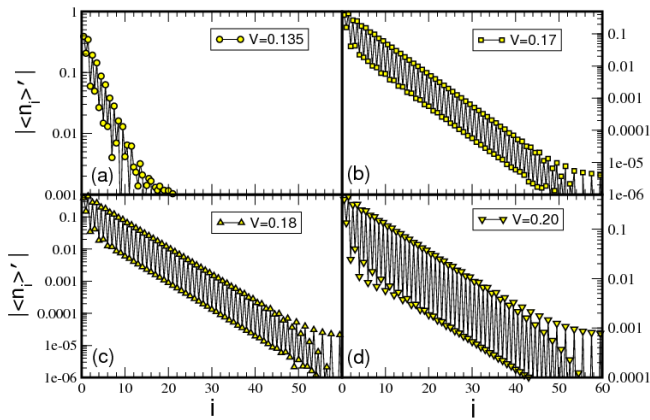}}
\caption{(Color online) The absolute value of the shifted density
  profile for the left half of the system at four values of $V$. The
  semi-log plot shows that the envelope of the edge states is
  exponential and that the exponent decreases as the HI-CDW transition
  is approached.}
\label{HIdenprof-shiftedlog}
\end{figure}

The envelopes of the edge states decay exponentially as is shown in
Fig.\ref{HIdenprof-shiftedlog}. The exponent decreases as $V$
increases toward the HI-CDW transition and the edge states penetrate
deeper into the system.

\section{Entanglement spectrum and Wannier functions }
\label{wanniersection}

\begin{figure}[h!]
\centerline{\includegraphics[width=9cm]{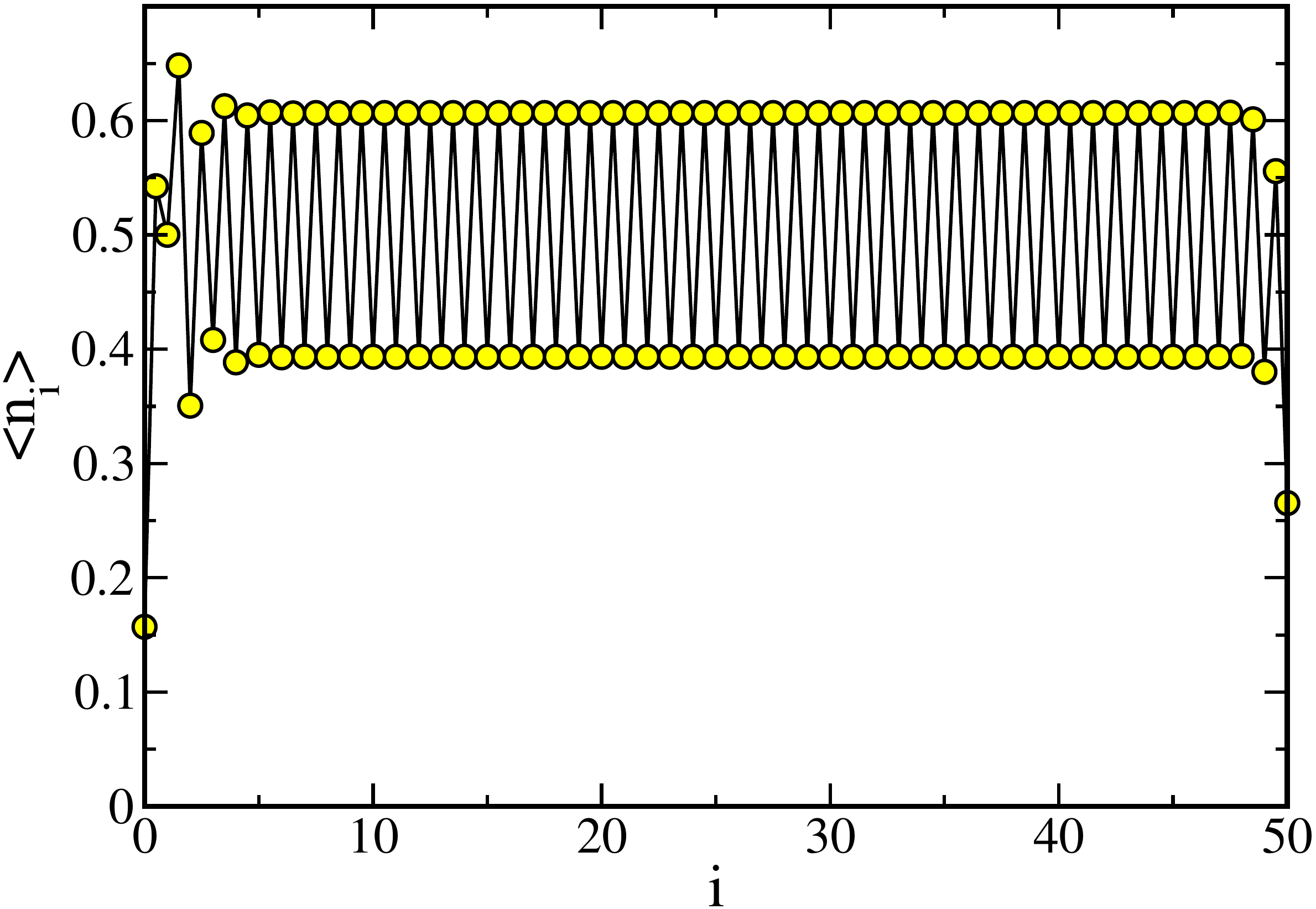}}
\caption{(Color online) Density profile of the AKLT-like state
  $|\Psi_{--}\rangle$.  Away from the boundaries, the average density
  on the $A$ site is 0.61 and 0.39 on the $B$ site; these values are
  extremely close to the DMRG ground state ones in the HI phase.  In
  addition, the edge states contain a fractional number of bosons, the
  imbalance being $\pm0.2$ like for the ground state of our model.}
\label{AKLT-wannier-state}
\end{figure}

Since the density profiles and the gaps agree very well with a
Haldane-like topological phase, we have also looked at the properties
of the entanglement spectrum in the different phases using DMRG and
imaginary time TEBD.  Interestingly, and contrary to the usual EBH
model, we found no degeneracy in the entanglement spectrum. At first
glance, the absence of degeneracy may appear puzzling. However, it
turns out that this feature does not mean that the phase is not
topological; in fact it is the consequence of the non-locality of the
Wannier state. Indeed it also appears in the AKLT-like matrix product
state built on the Wannier states rather than the local states. More
precisely, we consider the following family of states:
\begin{equation}
\left[\begin{array}{cc}
|\Psi_{-+}\rangle & |\Psi_{++}\rangle \\
|\Psi_{--}\rangle & |\Psi_{+-}\rangle\end{array}
\right]=\otimes_{i=1}^L\left[\begin{array}{cc}
\cos\theta|1,i\rangle & \sin\theta|2,i\rangle \\
\sin\theta|0,i\rangle & \cos\theta|1,i\rangle\end{array}
\right],
\end{equation}
where $|n\,i\rangle$ denotes the Fock state having $n$ bosons in the
Wannier state centered around the $B$ site $i$. Each of the four
preceding states has unit average density and a doubly degenerate
entanglement spectrum when computing the density matrix in the Wannier
basis. Expanding the Wannier states in the local states (see
Eq.~\eqref{wfb}), one can, in principle, compute the expression of
$|\Psi_{\pm\,\pm}\rangle$ in the local Fock states:
\begin{equation}
\label{akltexp}
 |\Psi_{\pm\,\pm}\rangle = \sum_{\{n_{A},n_B\}}
 C^{\pm\,\pm}_{\{n_{A},n_B\}} |\{n_{A},n_B\}\rangle, 
\end{equation}
where $\{n_{A},n_B\}$ is a shorthand notation for a particular
configuration of the occupation numbers in the different sawtooth
lattice sites, i.e.  $(\cdots,n_{A_i},n_{B_i},\cdots)$, and
$|\{n_{A},n_B\}\rangle$ is the Fock state corresponding to this
configuration.  Actually, since the number of coefficients grows
exponentially, exact numerical values can only be obtained for rather
small sizes $L\approx 10$. On the other hand, it is very simple to
build the matrix product state (MPS) with fixed bond dimension (in the
local Fock states $|\{n_{A},n_B\}\rangle$) associated with these
AKLT-like states. We have verified that for small system size, the MPS
approximation has the same properties (density) as the exact
expression given by Eq.~\eqref{akltexp}.  For the values
$\cos\theta=1/\sqrt{3}$ and $\sin(\theta)=\sqrt{2/3}$, the density
profile of the state $|\Psi_{--}\rangle$ is given in
Fig.~\ref{AKLT-wannier-state} for $L=50$.  Away from the boundaries,
the average density on the $A$ site is 0.61 and 0.39 on the $B$ site;
these values are extremely close to the DMRG ground state ones in the
HI phase, see Fig.~\ref{HIdenprof}.  More saliently, we find that the
entanglement spectrum obtained by splitting the state in the left and
right parts in the local basis is uniform and the largest eigenvalues
are: $(0.69, 0.24, 0.051, 0.014)$. Obviously, the spectrum is not
degenerate which emphasizes the impact of the non-locality of the
Wannier function on the properties of the entanglement. In addition,
the values found for the AKLT-like state are extremely close to the
ground state ones in the HI phase $(0.69 0.23 0.046 0.02)$, which
indirectly proves that for $U=0.5$ and $0.13<V<0.256$, the system does
exhibit a Haldane phase, but built on the Wannier states. Finally, one
can compute the number of bosons of the left and right edge states of
these AKLT states, as in in section~\ref{edgestatesection} for the
ground state. Here again, one finds that the left part has 25.2
particles and the right side has 24.8 particles, the imbalance is $\pm
0.2$ bosons, exactly as for the DMRG ground state.

\begin{figure}[h!]
\centerline{\includegraphics[width=9cm]{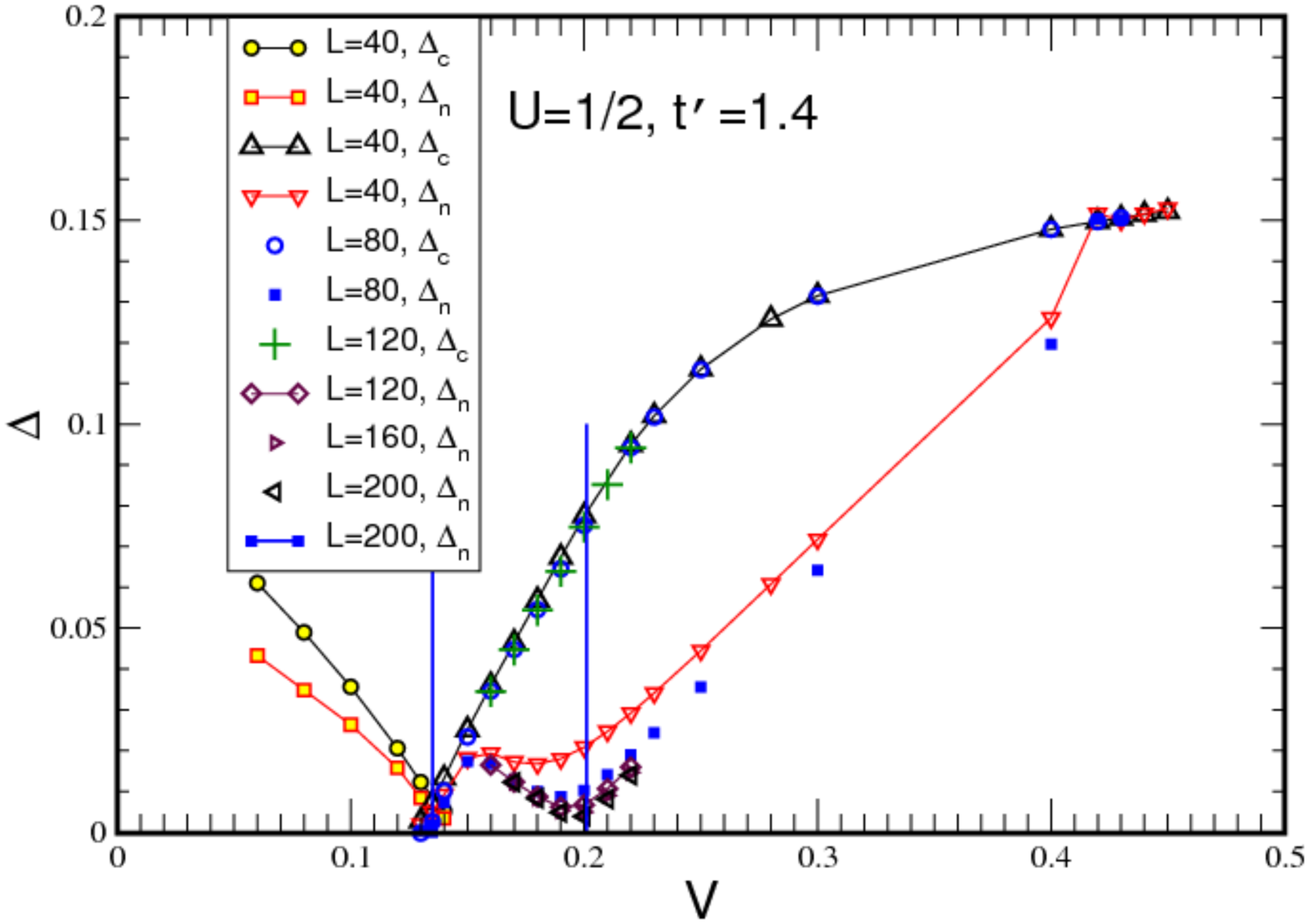}}
\caption{(Color online) DMRG results for the charge, $\Delta_c$, and
  neutral, $\Delta_n$, gaps as functions of the near neighbor
  repulsion, $V$, for $U=1/2$, $t^{\prime}=1.4t$ at half filling. The
  vertical lines show the locations of the MI-HI ($V\approx 0.136$)
  and the HI-CDW ($V=0.2$) transitions.}
\label{gapsU0.5t1.4}
\end{figure}

\begin{figure}[h!]
\centerline{\includegraphics[width=9cm]{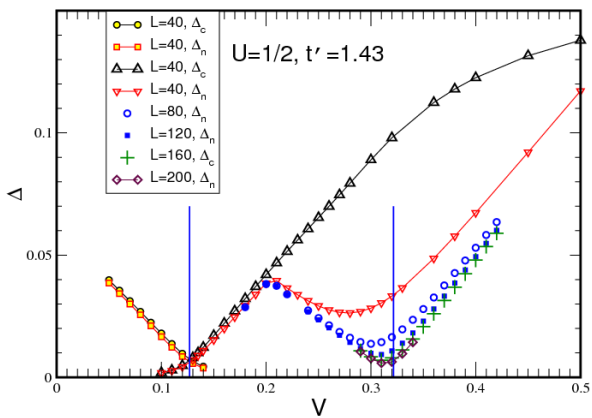}}
\caption{(Color online) DMRG results for the charge, $\Delta_c$, and
  neutral, $\Delta_n$, gaps as functions of the near neighbor
  repulsion, $V$, for $U=1/2$, $t^{\prime}=1.43t$ at half filling. The
  vertical lines show the locations of the MI-HI ($V\approx 0.128$) and
  the HI-CDW ($V=0.322$) transitions.}
\label{gapsU0.5t1.43}
\end{figure}

\section{Nearly flat band}
\label{notflatsection}

So far, we have studied the system only in the flat band case,
$t^{\prime}=\sqrt{2}t$. The question we address now is: Does the phase
diagram, in particular the HI phase, survive for any value of
$t^{\prime}$? We studied in detail the phase diagram for two values,
$t^{\prime}=1.4$ ({\it i.e.} smaller than $\sqrt{2}$) and
$t^{\prime}=1.43$ ({\it i.e.} larger than $\sqrt{2}$). The results,
Figs.\ref{gapsU0.5t1.4} and \ref{gapsU0.5t1.43}, show that the gaps
behave in a very similar manner to the $t^{\prime}=\sqrt{2}$ case, in
particular the behavior of $\Delta_n$ indicates that the HI phase
persists. Furthermore, we verified that the edge states are present
and behave as for the case of $t^{\prime}=\sqrt{2}t$.

However, it is clear from Fig.\ref{gapsU0.5t1.4} and
Fig.\ref{gapsU0.5} that the HI phase gets narrower with the smaller
value of $t^{\prime}$. In fact, the HI phase does not exist for
$t^{\prime}=1.3$. On the other hand, Fig.\ref{gapsU0.5t1.43} shows
that the HI phase expands for the larger value of
$t^{\prime}$. However, this expansion does not continue as
$t^{\prime}$ continues to increase; for a large enough value
$\Delta_n$ will not dip down to zero. The CDW phase will be reached
directly from the MI phase. This is the case for $t^{\prime}=1.5$.

This shows that the phase diagram, in particular the presence of the
topological HI phase, is robust and persists in a finite (but narrow)
range of values of $t^\prime$ centered at the flat band value,
$t=\sqrt{2}$. We also observed the same behavior for $U=1$.

\section{Conclusion and outlook}
\label{conclusionsection}

In summary, we have studied the phase diagram of the half filled
Bose-Hubbard system in the sawtooth lattice in the situation where the
geometric frustration in the hopping term produces a flat band. We
have shown that, in the presence of contact and near neighbor
repulsion, three phases exist: Mott insulator (MI), charge density
wave (CDW), and the topological Haldane insulating (HI) phase. In
particular, we have shown that in the HI phase, even though the
entanglement spectrum is not doubly degenerate, it is in excellent
agreement with the entanglement spectrum of the
Affleck-Kennedy-Lieb-Tasaki (AKLT) state built in the Wannier basis
associated with the flat band.  This emphasizes that the abscence of
degeneracy in the entanglement spectrum is not necessarily a
signature of a non-topological phase, but rather that the (hidden)
protecting symmetry involves non-local states.  Finally, we have also
shown that, at fixed interactions, the HI phase is stable against
small departure from flatness of the band but is destroyed as the band
dispersion becomes stronger.
  
For future work, it would be interesting to find an efficient way to
compute the string order in the Wannier basis and to show that it
vanishes at the MI-HI transition. In addition, it would be
illuminating to study the excitations of the system, and, along the
preceding line, to show that the elementary excitations correspond to
domain walls in the string order. 
  
\begin{acknowledgments}
  This research is supported by the National Research Foundation,
  Prime Minister’s Office, Singapore and the Ministry of Education,
  Singapore under the Research Centres of Excellence programme.
\end{acknowledgments}

\end{document}